\documentclass[useAMS,usenatbib,fleqn,a4paper]{mn2e}
\usepackage{times}
\usepackage{amsmath}
\usepackage{amssymb}
\usepackage{amsbsy}
\usepackage{bmpsize}
\usepackage{graphicx}
\usepackage{subfigure}
\usepackage{paralist}
\usepackage{mathrsfs}
\usepackage{booktabs}
\usepackage{tabularx}
\usepackage{courier}
\usepackage{verbatim}

\def\df{{\sc df}}

\def\Rc{R_{\rm c}}

\def\Ra{R_{\rm a}}
\def\Rp{R_{\rm p}}
\def\ra{r_{\rm a}}
\def\rp{r_{\rm p}}

\def\kpc{\,{\rm kpc}}

\def\pc{\,{\rm pc}}

\def\kms{\,{\rm km\,s^{-1}}}

\def\percent{\text{ per cent}}
\def\half{{\textstyle{\frac12}}}

\newcommand{\bs}[1]{\bmath{#1}}

\def\percent{\text{ per cent}}

\newcommand{\vJ}{\bs{J}}\newcommand{\vx}{\bs{x}}
\newcommand{\vv}{\bs{v}}

\usepackage{color}
\definecolor{darkred}{rgb}{0.55, 0.0, 0.0}
\definecolor{darkblue}{rgb}{0.0, 0.0, 0.55}
 \definecolor{darkgreen}{rgb}{0.0, 0.2, 0.13}
\usepackage[colorlinks=true,linkcolor=darkred,citecolor=darkblue,urlcolor=darkgreen]{hyperref}
\title[Action estimation methods for galactic dynamics]{A review of action estimation methods for galactic dynamics}
\author[J. L. Sanders \& J. Binney]{Jason L. Sanders$^1$\thanks{E-mail: jls@ast.cam.ac.uk} \& James Binney$^2$\\
$^1$Institute of Astronomy, Madingley Road, Cambridge, CB3 0HA\\
$^2$Rudolf Peierls Centre for Theoretical Physics, Keble Road, Oxford, OX1 3NP, UK}

\date{Accepted XXX. Received YYY; in original form ZZZ}

\begin{document}
\label{firstpage}
\pagerange{\pageref{firstpage}--\pageref{lastpage}} \pubyear{2015}
\maketitle
\begin{abstract}
We review the available methods for estimating actions, angles and
frequencies of orbits in both axisymmetric and triaxial potentials. The
methods are separated into two classes. Unless an orbit has been trapped by a
resonance, convergent, or iterative, methods are able to recover the actions
to arbitrarily high accuracy given sufficient computing time.  Faster
non-convergent methods rely on the potential being sufficiently close to a
separable potential and the accuracy of the action estimate cannot be
improved through further computation.  We critically compare the accuracy of
the methods and the required computation time for a range of orbits in an
axisymmetric multi-component Galactic potential.  We introduce a new method
for estimating actions that builds on the adiabatic approximation of
Sch\"onrich \& Binney (2012) and discuss the accuracy required for the actions, angles and frequencies using suitable distribution functions for the
thin and thick discs, the stellar halo and a star stream. We conclude that for studies of the disc and smooth halo component of the Milky Way the most suitable compromise between speed and accuracy is the St\"ackel Fudge, whilst when studying streams the non-convergent methods do not offer sufficient accuracy and the most suitable method is computing the actions from an orbit integration via a generating function. All the software used in this study
can be downloaded from
\href{https://github.com/jls713/tact}{https://github.com/jls713/tact}.
\end{abstract}


\section{Introduction}

Galaxies are complex dynamical systems in which stars and dark-matter
particles move on orbits in the gravitational field that is generated by all
the other stars and dark-matter particles. Early experiments with typical
smooth, axisymmetric galactic potentials demonstrated that these orbits
possess three integrals of motion \citep{Ollongren1962}. By Jeans' theorem,
the distribution function (DF) of an equilibrium galaxy model can be assumed to be
a function of these integrals of motion.  Any function of the integrals of
motion is also an integral but action integrals stand out from the rest
because only action integrals $J_i$ can be complemented by canonically
conjugate variables $\theta_i$ to form a complete system $(\btheta,\bs{J})$ of
canonical coordinates. The actions $\bs{J}$ label orbits and each angle
variable $\theta_i$ increases linearly in time at the rate $\Omega_i$.  For a
discussion of the merits of these variables we refer readers to
\cite{BinneyTremaine}.

It has been demonstrated that DFs that are analytic functions of the actions
can successfully model near-equilibrium structures in our Galaxy such as the
disc \citep{Binney2010,Binney2012b,BovyRix2013,Piffl2014,SandersBinneyEDF},
the stellar halo \citep[][Das \& Binney, in prep.]{WilliamsEvans2015} and the dark halo
\citep{BinneyPiffl2015}.  Moreover, angle-action variables have proved useful for
modelling structures that are not phase-mixed, such as tidal streams
\citep{HelmiWhite1999,SandersBinney2013a,Bovy2014,Sanders2014} and
substructures in the velocity space of solar-neighbourhood stars
\citep{Sellwood2010,McMillan2011}.

Although orbits in typical potentials appear to admit three action integrals,
the actions can be computed analytically only in a very few cases. Hence
practical use can be made of angle-action coordinates only to the extent that
numerical methods enable us to compute angle-action coordinates from normal
$(\bs{x},\bs{v})$ coordinates . In recent years, and with the arrival of the
\emph{Gaia} data on the horizon, much effort has been invested in developing
such methods. Many of these methods rely on, or are inspired by, the rare
cases in which we can compute the actions analytically.  This paper
summarises and collates these methods, and critically compares the various
approaches.  We seek to guide and advise readers on the best methods for
approaching different types of data. In addition we have made available code
for all the approaches detailed in this paper at
\href{https://github.com/jls713/tact}{https://github.com/jls713/tact}.

The paper is organised as follows. In Section~\ref{Sect::Separable} we recall
the cases in which the actions can be computed exactly. We then describe
numerical methods for generic potentials under two headings: ``convergent''
methods in Section~\ref{Sect::MethodsC} and ``non-convergent'' methods in
Section~\ref{Sect::MethodsNC}. One of these methods, the Spheroidal Adiabatic
Approximation (SAA), has not been previously published. In
Section~\ref{Sect::MethodComparison} we critically compare the methods under
two headings: accuracy and computational cost. In Section \ref{Sect::Discuss}
we compare the accuracy required in various applications to the accuracy
achieved by the methods that deliver best value. The use of the presented methods in non-integrable Hamiltonians and the impact of resonant trapping is discussed. We also
briefly discuss how changes of parameters affect the accuracy of one of the
most powerful methods, and we outline the structure of the code we have made
available. Section \ref{Sect::Conclusions} sums up. An appendix gives
functional forms of DFs that provide good models of the thin and thick discs, the stellar halo
and a stellar stream.

\section{Separable potentials}\label{Sect::Separable}

Angle-action coordinates are intimately linked to separable potentials. A
separable potential is one for which the Hamilton-Jacobi equation can be
solved by separation of variables. The separation process introduces two
constants of motion $I_2$ and $I_3$ in addition to the energy $E$ and yields an
expression for each momentum $p_i(q_i;E,I_2,I_3)$ as a function of one
coordinate $q_i$ and the integrals of motion. Given these expressions, the
actions $J_i=(2\pi)^{-1}\oint\mathrm{d} q_i\,p_i$ can be computed as
functions $J_i(E,I_2,I_3)$ of the ``classical'' integrals of motion.

\subsection{Spherical potentials}

The Hamilton-Jacobi equation for any spherical potential, $\Phi(r)$, is
always separable. The Hamiltonian for a particle in a spherical potential is
\begin{equation}
H = \half p_r^2 +\frac{L^2}{2r^2} + \Phi(r),
\label{SphericalH}
\end{equation}
 where $L=|\bs{L}|$ is the length of the angular momentum vector.  The orbit
is confined to the plane normal to $\bs{L}$, and in this plane the angular
motion separates from the radial motion. The angular motion is quantified by
the angular action $L$. The
Hamiltonian~\eqref{SphericalH} is one-dimensional, so the radial
action is
\begin{equation}\label{JrSphere}
J_r =
\frac{1}{\pi}\int_{\rp}^{\ra}\mathrm{d}r\sqrt{2E-2\Phi-\frac{L^2}{r^2}},
\end{equation}
where $\rp$ is the pericentric radius and $\ra$ the
apocentric radius.

In a spherical potential the third action is  arbitrary in the sense that it
defines the orientation of the orbital plane with respect to some coordinate
system.  It is convenient to take it to be $J_\phi=L_z$ i.e., the component
of angular momentum along the $z$-axis. We have the freedom to use as new
actions linear combinations of old actions and we use this freedom to use the
set $\bs{J}=(J_r,L_z,L-|L_z|)$, which proves to be the analogue of the sets
we will work with in the  axisymmetric but non-spherical case.

\subsubsection{Analytic cases}\label{Sect::analytic}

The integral (\ref{JrSphere}) can rarely be evaluated analytically.
An exception of great importance  is the isochrone
potential \citep{Henon}
\begin{equation}
\Phi(r) = -\frac{GM}{b+\sqrt{r^2+b^2}},
\end{equation}
 where $M$ is the mass and $b$ is a scale radius. In this potential the
actions and angles are analytic functions of $(\bs{x},\bs{v})$. In the limit
$b\rightarrow0$ the isochrone tends to the Kepler potential, whilst in the
limit $b\rightarrow\infty$ the isochrone tends to the spherical
harmonic-oscillator potential.  More generally, we have analytic expressions
for angle-action coordinates as functions of Cartesian phase-space coordinates
for any triaxial harmonic-oscillator potential
\begin{equation}
\Phi(\bs{x}) = \frac{1}{2}\sum_{i=1}^3\omega_i^2{x_i}^2,
\end{equation}
where $\omega_i$ are the oscillator's frequencies.

\subsection{St\"ackel potentials}\label{StackelPot}

The most general class of separable potentials is that of triaxial St\"ackel
potentials. Below we will discuss the subclass of axisymmetric St\"ackel
potentials, and spherical potentials may be considered to lie within this
subclass. Triaxial St\"ackel potentials are intrinsically linked to confocal
ellipsoidal coordinates. Here we briefly detail the relevant properties of
these coordinates.

\subsubsection{Confocal ellipsoidal coordinates}

The confocal ellipsoidal coordinates $(\lambda,\mu,\nu)$ of the point with
Cartesian coordinates $(x,y,z)$ are defined to be the three roots
$\nu\le\mu\le\lambda$ of the
cubic in $\tau$
\begin{equation}
\frac{x^2}{(\tau-a^2)}+\frac{y^2}{(\tau-b^2)}+\frac{z^2}{(\tau-c^2)} = 1.
\end{equation}
 Here $a$, $b$ and $c$ are constants that define the coordinate system. We
adopt the convention that the $x$ axis is the potential's major axis,
$y$ is the intermediate axis and $z$ is the minor axis. For this to be the
case we require
$c^2\leq\nu\leq b^2\leq\mu\leq a^2\leq\lambda$. Surfaces of constant
$\lambda$ are ellipsoids, surfaces of constant $\mu$ are hyperboloids of one
sheet (flared tubes of elliptical cross section that surround the $x$ axis),
and surfaces of constant $\nu$ are hyperboloids of two sheets that have their
extremal points on the $z$ axis. In the plane $z=0$, lines of constant
$\lambda$ are ellipses with foci at $y=\pm\Delta_1\equiv\pm\sqrt{a^2-b^2}$,
whilst in the plane $x=0$ lines of constant $\mu$ are ellipses with foci at
$z=\pm\Delta_2\equiv\pm\sqrt{b^2-c^2}$. Adding the same constant to $a^2$,
$b^2$ and $c^2$ simply adds that constant to each of $\lambda$, $\mu$ and
$\nu$, leaving invariant $\Delta_1$ and $\Delta_2$ and the shapes of the
coordinate curves. One may exploit this degeneracy to set one of $a$, $b$ or
$c$ to zero.

 The generating function, $S$ of the canonical transformation between Cartesian,
$(x,y,z,p_x,p_y,p_z)$, and ellipsoidal coordinates,
$(\lambda,\mu,\nu,p_\lambda,p_\mu,p_\nu)$ is
\begin{equation}
S(p_x,p_y,p_z,\lambda,\mu,\nu) = p_x x(\lambda,\mu,\nu)
+p_y y(\lambda,\mu,\nu)
+p_z z(\lambda,\mu,\nu).
\end{equation}
The momentum conjugate to any ellipsoidal coordinate $\tau$ is $p_\tau =
\upartial S/\upartial \tau$.

When we express the Hamiltonian as a function of
ellipsoidal coordinates, we find
\begin{equation}
\begin{split}
H &= \half(p_x^2+p_y^2+p_z^2)+\Phi(x,y,z)
\\&=\half\Big(\frac{p_\lambda^2}{P_\lambda^2}+\frac{p_\mu^2}{P_\mu^2}+\frac{p_\nu^2}{P_\nu^2}\Big)+\Phi(\lambda,\mu,\nu).
\end{split}
\label{Eq::Hamiltonian}
\end{equation}
where
\begin{equation}
P^2_\lambda = \frac{(\lambda-\mu)(\lambda-\nu)}{4(\lambda-a^2)(\lambda-b^2)(\lambda-c^2)},
\end{equation}
and $P_\mu$ and $P_\nu$ are given by cyclic permutations of $(\lambda,\mu,\nu)$.

\subsubsection{Triaxial St\"ackel potentials}

The most general triaxial St\"ackel potential is
\begin{equation}
\Phi_{\rm S}(\lambda,\mu,\nu) = \frac{F(\lambda)}{(\lambda-\mu)(\nu-\lambda)}+\frac{F(\mu)}{(\mu-\nu)(\lambda-\mu)}+\frac{F(\nu)}{(\nu-\lambda)(\mu-\nu)}.
\end{equation}
$\Phi_{\rm S}$ is composed of three functions of one variable. Here we denote the
three functions with the same letter, $F$, as their domains are distinct.
Moreover, for $\Phi_{\rm S}$ to be finite at $\lambda=\mu=a^2$ and
$\mu=\nu=b^2$, $F(\tau)$ must be continuous at $\tau=a^2$ and $\tau=b^2$.
Solution of  the Hamilton-Jacobi equation
\citep{deZeeuw1985a} yields equations for the momenta of the form
\begin{equation}
2(\tau-a^2)(\tau-b^2)(\tau-c^2)p_\tau^2=\tau^2 E -\tau A+B + F(\tau),
\label{Eq::EqnOfMotion}
\end{equation}
 where $A$ and $B$ are separation constants.  Given a phase-space
point, $(\vx_0,\vv_0)$, we find $\tau_0(\vx_0,\vv_0)$ and $p_{\tau
0}(\vx_0,\vv_0)$ from  the coordinate transformations, and can then find the
integrals $A$ and $B$ by solving equation~\eqref{Eq::EqnOfMotion} simultaneously
for two coordinates, say $\tau=\mu$ and $\tau=\lambda$ \citep[see][for more details]{deZeeuw1985a}. These integrals are related to
the classical integrals $I_2$ and $I_3$ in a simple way:
\begin{equation}
\begin{split}
I_2&=\frac{a^4 E-a^2 A+B}{c^2-a^2},\\
I_3&=\frac{c^4E-c^2 A+B}{a^2-c^2}.
\end{split}
\end{equation}
As $p_\tau$ is a function of  only $\tau$, the actions are then given by the
1D integrals
\begin{equation}
J_\tau = \frac{2}{\upi}\int_{\tau_-}^{\tau_+} \mathrm{d}\tau\,|p_\tau(\tau)|.
\label{Eq::action}
\end{equation}
where ($\tau_-,\tau_+$) are the roots of $p_\tau(\tau)=0$, which we locate
with Brent's method. We compute the integral (and all similar integrals) using a $10$-point Gauss-Legendre
quadrature. \cite{deZeeuw1985a} chose to define the actions in a triaxial St\"ackel potential in this way such that the actions are continuous across the transitions from one orbit family to the next. However, for the loop orbits the oscillation in one of the coordinates is covered twice. In the axisymmetric case this means that the radial action $J_\lambda$ is half the value given by the right-hand side of equation~\eqref{Eq::action}.

\subsubsection{Axisymmetric limits}

In the limit $a^2\rightarrow b^2$ the confocal ellipsoidal coordinate system
reduces to a prolate spheroidal coordinate system. In this case $\Delta_1\rightarrow0$ and the $\mu$
coordinate is replaced by the familiar polar angle $\phi$. The coordinate system is now described by a single pair of foci separated by $\Delta=\Delta_2=\sqrt{a^2-c^2}$. The most general
potentials that are separable  in these coordinates, are the oblate St\"ackel potentials
\begin{equation}
\Phi_{\rm S,obl} = -\frac{F(\lambda)-F(\nu)}{\lambda-\nu}.
\label{eqn::axi_stackel}
\end{equation}
 The potential is defined by two functions of a single variable $F(\tau)$
that have distinct domains. In these potentials $I_2=\frac{1}{2}J_\phi^2$.
Note that taking the limit $b^2 \rightarrow c^2$ reduces the confocal
ellipsoidal coordinate system to an oblate spheroidal coordinate
system and the corresponding separable potentials are the prolate St\"ackel
potentials. There is convincing evidence that the potential of the Milky Way
is oblate, so we will not discuss prolate potentials.

In the axisymmetric case,
equation~\eqref{Eq::EqnOfMotion} yields $p_\tau(\tau)$ in the form
\begin{equation}
2(\tau-a^2)(\tau-c^2)p_\tau^2=(\tau-c^2)E-\frac{(\tau-c^2)}{(\tau-a^2)}I_2-I_3+F(\tau).
\label{Eq::EqnOfMotionAxi}
\end{equation}

From equation~\eqref{eqn::axi_stackel} we see that the derivative
\[
\frac{\partial^2}{\partial\lambda\partial\nu}\Big[(\lambda-\nu)\Phi_{\rm S,obl}\Big]=0.
\]
This expression can be rewritten in terms of $R$ and $z$ derivatives
as\footnote{This equation corresponds to equation (8) of \cite{Sanders2012a} except in that paper a minus sign is
missing on the right hand side of the equation.}
\begin{equation}
\begin{split}
\Delta^2 &= a^2-c^2 \\
&= z^2-R^2+\Big[3z\frac{\partial \Phi}{\partial R}
-3R\frac{\partial \Phi}{\partial z}\\
&\quad+Rz\Big(\frac{\partial^2 \Phi}{\partial R^2}-\frac{\partial^2 \Phi}{\partial
z^2}\Big)\Big]\bigg/\frac{\partial^2 \Phi}{\partial R\partial z},
\end{split}
\label{DeltaGuess}
\end{equation}
where we have dropped the subscript S,obl for brevity. We will use this expression to find an approximate $\Delta$ for general axisymmetric potentials in Section~\ref{Sect::MethodsNC}.

\section{Convergent methods}\label{Sect::MethodsC}

In this and the following section we present the available methods for
estimating actions.  We present each method briefly since full details can be
found in cited papers. Whenever numerical integration of the equations of
motion is required, we use an adaptive embedded Runge-Kutta Prince-Dortmund
(8,9) scheme provided in the Gnu Science Library \citep{GSL} with a relative
accuracy of $10^{-8}$.

We split methods for estimating actions into two categories:
\emph{convergent} and \emph{non-convergent}. In the absence of resonant
trapping, convergent methods are such that the errors in the computed actions
can be reduced to arbitrarily small values given sufficient computation.
Non-convergent methods, by contrast, rely on an approximation that cannot be
refined, so the error in the computed actions cannot be arbitrarily
diminished. In this section we discuss convergent methods.

These methods all involve numerical construction of the generating function that maps
the analytic angle-action variables $(\btheta',\boldsymbol{J}')$ of a ``toy''
potential such as the isochrone or harmonic oscillator, into the angle-action
variables $(\btheta,\boldsymbol{J})$ of the true potential.

\subsection{Iterative torus construction (ItTC)}\label{Method::ItTorus}

\cite{McGillBinney} introduced the technique of ``torus mapping'' to obtain
actions for a general axisymmetric potential. They used
the isochrone potential as the toy potential, and showed that the required
generating function can be written
\begin{equation}\label{eq:genGF}
S(\btheta',\bs{J}) = \btheta'\cdot\bs{J}
+2 \sum_{\bs{n}>\bs{0}}S_{\bs{n}}(\bs{J})\sin\bs{n}\cdot\btheta',
\end{equation}
where the vector $\bs{n}$ has integer components: on account of the
periodicity of the angle coordinates, the generating function is a Fourier
series with coefficients $S_{\bs{n}}$. On account of the time-reversal
symmetry of the Hamiltonian and assumed mirror symmetry of the potential
around the plane $z=0$, we require only a sine series and only even values of
$n_z$ occur. From the oddness of $\sin(x)$, the sum over $\bs{n}$ can be
restricted to only half of $\bs{n}$ space.

The relation between the toy and true
actions is
\begin{equation}
\bs{J}'=\frac{\partial S}{\partial \btheta'} = \bs{J}+2\sum_{\bs{n}>\bs{0}}\bs{n}S_{\bs{n}}(\bs{J})\cos\bs{n}\cdot\btheta',
\label{toyact}
\end{equation}
while the angles are related by
\begin{equation}
\btheta=\frac{\partial S}{\partial \bs{J}}
= \btheta'+2\sum_{\bs{n}>\bs{0}}\frac{\partial S_{\bs{n}}}{\partial \bs{J}}(\bs{J})\sin\bs{n}\cdot\btheta'.
\label{targetang}
\end{equation}
 Given a true action $\bs{J}$ and a trial set of coefficients
$S_{\bs{n}}$, for each point on a regular grid in toy angle $\btheta'$ the
corresponding toy actions $\bs{J}'$ are found from equation~\eqref{toyact}.
The coordinates $(\bs{x},\bs{v})$ are then recovered from the toy
potential's analytic formulae, and the true Hamiltonian is evaluated at
this phase-space point.  The Marquardt-Levenberg algorithm is used to adjust
the $S_{\bs{n}}$ iteratively until the spread in energy at the sampled points
is minimised. \cite{BinneyKumar} showed how the procedure could be extended
to recover $\partial S_{\bs{n}}/\partial\bs{J}$ so the true angles can be
recovered from equation \eqref{targetang}.  \cite{KaasalainenB} showed that
torus mapping can be also used for potentials that admit more than one orbit
family and introduced a new algorithm for the recovery of $\partial
S_{\bs{n}}/\partial\bs{J}$.

This algorithm involves integrating short sections of the orbit starting from
several points on the torus. For each point along each orbit section, the toy
angles $\btheta'$ are known and the true angles $\btheta$ must satisfy
$\btheta=\btheta(0)+\boldsymbol{\Omega}t$. Equating these values to those
given by equation \eqref{targetang}, we obtain a series of
linear equations for $\btheta(0)$, $\boldsymbol{\Omega}$ and $\partial
S_{\bs{n}}/\partial\bs{J}$:
\begin{equation}
\btheta(0)+\boldsymbol{\Omega}t
= \btheta'+2\sum_{\bs{n}>\bs{0}}\frac{\partial S_{\bs{n}}}{\partial \bs{J}}(\bs{J})\sin\bs{n}\cdot\btheta'.
\label{toyang}
\end{equation}
 More equations are obtained than one has unknowns and they are solved in a
least-squares sense by standard methods.

With the $S_{\bs{n}}$ and $\partial S_{\bs{n}}/\partial \bs{J}$ found, the
corresponding $(\bs{x},\bs{v})$ for any point $\btheta$ on the torus
$\boldsymbol{J}$ can be found. This is ideal for many applications but for
others one needs machinery to convert $(\bs{x},\bs{v})$ to
$(\btheta,\boldsymbol{J})$ \citep[e.g.][]{McMillanBinney2013}.
\cite{McMillanBinney2008} obtained $(\btheta,\boldsymbol{J})$ for given
points $(\bs{x},\bs{v})$ by repeatedly constructing tori until one was
obtained that passed through the given phase-space point. Their procedure was
slow but \cite{SandersBinney2015} proposed a much faster iterative
procedure that combines torus mapping with one of the non-convergent methods
described below. We call this method \emph{Iterative Torus Construction} (ItTC).

One first estimates the angle-actions $(\btheta_{\rm S},\boldsymbol{J}_{\rm S})$ from
$(\bs{x},\bs{v})$ using the axisymmetric St\"ackel Fudge of
Section~\ref{Method::SF_Axi}. A torus with actions $\boldsymbol{J}_{\rm S}$ is
constructed and the point $(\bs{x}_{\rm S},\bs{v}_{\rm S})$ on this torus that is nearest to the
given point $(\bs{x},\bs{v})$ is found by minimization of the distance
\[
\eta=|\bs{\Omega}|^2(\bs{x}-\bs{x}_S)^2+(\bs{v}-\bs{v}_S)^2,
\]
with respect to the angles $\btheta$ using $\btheta_{\rm S}$ as an initial guess. $\bs{\Omega}$ is the frequency of the constructed torus. One now uses the
St\"ackel Fudge to estimate the angle-actions $(\btheta_P,\boldsymbol{J}_P)$
of the point $(\bs{x}_{\rm S},\bs{v}_{\rm S})$. One now has in
$\Delta\bs{J}=\bs{J}_P-\bs{J}_{\rm S}$ an estimate of the error in the actions
produced by the St\"ackel Fudge.  Under the
assumption that this error varies slowly in phase space, one's new estimate
of the action is then
$\boldsymbol{J}_{\rm S}-\Delta\boldsymbol{J}=2\boldsymbol{J}_{\rm S}-\boldsymbol{J}_P$.

The procedure can be repeated using this improved action estimate to
construct another torus. \cite{SandersBinney2015} found that a single torus
construction produced actions accurate to $0.01\percent$ and further torus
constructions only reduced this by a factor of two. We decide that the algorithm has converged when the distance $\eta$ falls below some threshold $\eta^*$. For our standard setup we
use $\eta^*=(0.1\kms)^2$, a maximum of $5$ torus constructions and a relative
error for each torus construction of $\Delta J/J=10^{-3}$.

\subsection{Generating function from orbit integration (O2GF)}\label{Method::Genfunc}

\cite{SandersBinney2014} proposed a method for constructing the generating
function \eqref{eq:genGF} that is based on orbit integration rather than
Marquardt-Levenberg minimisation of the variance in the Hamiltonian over a
trial torus. For brevity we will refer to this as the O2GF method. One starts
by computing $N_\mathrm{samp}$ phase-space points along an orbit integrated
for a time $N_T T$, where $T$ is the period of a circular orbit with the same
energy. At each phase-space point the actions and angles in a toy potential
are computed. These toy angle-actions are used to set up the linear equations
\eqref{toyact} and \eqref{toyang}. Finally, the entire set of equations is
solved for the unknowns: the true actions $\bs{J}$, the Fourier
coefficients, $S_{\bs{n}}$, and their derivatives $\partial
S_{\bs{n}}/\partial\bs{J}$ plus $\btheta(0)$ and $\bs{\Omega}$. The
vectors included in the sums are limited to $\bs{n}<N_\mathrm{max}$.

In \cite{SandersBinney2014} the parameters of the toy potential were chosen
by minimising the deviation of the toy Hamiltonian around the orbit. This
procedure was computationally costly and sub-optimal in that it tended to
produce very sharp variations of the toy Hamiltonian along the orbit that
required a large number of Fourier coefficients to remove. Here we find the
minimum and maximum radius of the orbit sample and require the radial force
of the toy and true potentials to agree there. Since we only consider two toy
parameters (the scale mass and radius of the isochrone), this procedure is
sufficient.

Several criteria for a sufficient sampling of toy angle space were discussed
by \cite{SandersBinney2014}. For each vector $\bs{n}$ we compute the minimum
and maximum values of $\bs{n}\cdot\btheta'$ (where $\btheta'$ are unrolled continuous
angles i.e. not $2\pi$ periodic). If the difference between the minimum and
maximum value of $\bs{n}\cdot\btheta'$ is less than $2\pi$ we repeat the algorithm
with $N_T\rightarrow 2N_T$. Secondly if for any mode
$\hbox{range}(\bs{n}\cdot\btheta')/N_T>\pi$ the density of the sampling is too low to
constrain the mode $\bs{n}$ and we repeat the algorithm with
$N_\mathrm{samp}\rightarrow 2N_\mathrm{samp}$. For our standard setup we use
$N_T=8$, $N_\mathrm{samp}=300$ and $N_{\rm max}=8$.

Using this approach actions can be found in a triaxial potential that
supports several orbital families \citep{SandersBinney2014}. A different toy
potential is used for each class of orbit, so the class (box, long-axis loop,
etc.) to which an orbit belongs must be established before the equations are
set up.  This can be done by determining whether the sign of the angular
momentum components $L_x$ and $L_z$ changes over the orbit. If both
components change sign, the orbit is a box orbit and a triaxial harmonic
oscillator potential is appropriate. If $L_z$ does not change sign, the orbit
is a short-axis loop and an isochrone potential with its symmetry axis along
$z$ is suitable. If $L_x$ does not change sign, the orbit is a long-axis loop
and an isochrone with symmetry axis along $x$ is suitable.

\subsection{Averaged generating function (AvGF)}\label{Method::AvGenfunc}

As noted by \cite{Bovy2014} and \cite{Fox2014}, averaging
equation~\eqref{toyact} over the toy angles yields an expression for the
true actions:
\begin{equation}
\bs{J}=\int{\mathrm{d}^3\btheta'\over(2\pi)^3}
\Big(\boldsymbol{J}'-2\sum_{\bs{n}>\bs{0}}\bs{n}S_{\bs{n}}(\bs{J})\cos\bs{n}\cdot\btheta'\Big)
=\int{\mathrm{d}^3\btheta'\over(2\pi)^3}\,\boldsymbol{J}'.
\end{equation}
 The task now is to estimate the value of the integral on the right given values
of $\bs{J}'$ at irregularly distributed points $\btheta'$. This can be done to
reasonable accuracy provided the points $\btheta'$ provide good coverage of
the basic cube of angle space. This ``AvGF'' method avoids solving elaborate matrix
equations, so it is simple to code. For the standard setup we use the same
parameters as for the O2GF method.

\section{Non-convergent methods}\label{Sect::MethodsNC}

The non-convergent methods below are all based on the simplicity with
which the actions can be computed in the separable potentials of
Section \ref{Sect::analytic}. The basic idea is to proceed as if the potential were of St\"ackel
form. As we require the actions for an individual orbit, it is not necessary
for a St\"ackel potential to provide a good global fit to the potential; it
suffices for a St\"ackel potential to provide a good fit over the region
explored by the orbit. For orbits that stay close to the plane, we will see that the
assumption of cylindrical separability works well, whilst for more vertically
and radially extended orbits a better approximation is found to be
separability in prolate spheroidal coordinates.  For even more radially and
vertically extended orbits the assumption of separability over the orbit
region breaks down, and convergent methods are required for decent accuracy.

The ideas presented here are connected to work on estimating a third
integral $I_3$ for general axisymmetric potentials. Any third integral can be
used as the argument of a distribution function although, as argued in the
introduction and elsewhere \citep[e.g.][]{BinneyMcMillan2015}, actions are
the integrals of choice from several perspectives. Two approaches to the
construction of $I_3$ have been tried: explicit fitting of a St\"ackel
potential to the real potential, or using the real potential in formulae
derived for St\"ackel potentials.

\cite{DejonghedeZeeuw1988} presented a general formalism for fitting a
St\"ackel potential to a general axisymmetric potential either globally or
locally, and \cite{BatsleerDejonghe} and \cite{Famaey2003} fitted a St\"ackel
potential to the available constraints for our Galaxy.  However, it seems the
inner regions of the Galaxy cannot be accurately represented with a St\"ackel
potential.  \cite{deBruyne2000} produced a series of locally-fitted St\"ackel
potentials and found that $I_3$ varied by $\sim10\percent$ around an orbit.

\cite{KentdeZeeuw1991} pioneered using the real potential in formulae
derived for St\"ackel potentials by considering the target potential along lines of constant prolate spheroidal coordinate. These authors proposed several methods for estimating $I_3$ based on this idea. For general disc orbits they found that the most accurate technique
was the `least-squares method' that minimised the variation in the expression
for $I_3$ in a St\"ackel potential using an approximation for
$F(\lambda)\approx-(\lambda-c^2)\Phi(R(\lambda,c^2),0)$.  More recently,
\cite{Bienayme2015} used what appears to be an identical approximation for
$F(\lambda)$ to that used by \cite{KentdeZeeuw1991} to demonstrate that a
third integral for disc orbits in the Besan\c con Galaxy model is conserved to
$\sim 1\percent$.

\subsection{Cylindrical adiabatic approximation (CAA)}\label{Method::PAA}

The cylindrical adiabatic approximation (CAA) was introduced by
\cite{Binney2010} to model the distribution of stars in the Galactic disc. He
argued that since the vertical frequency $\Omega_z$ of a disc star is
significantly larger than its radial frequency $\Omega_r$, the potential that
governs vertical oscillations may be considered to vary slowly as the star
oscillates radially, with the consequence that $J_z$ is adiabatically
invariant.  \citet{BinneyMcMillan2011} showed that a better approximation to
orbits can be obtained by replacing $J_\phi$ by $|J_\phi|+J_z$ in the formula
for the effective radial potential.  \citet{Schonrich2012} observed that
conservation of energy requires that variation along the orbit in the energy
$E_z$ of vertical motion is balanced by variation in the energy of radial
motion. This principle leads to a slightly different modification of the
effective radial potential to that proposed by \citet{BinneyMcMillan2011}.
Here we present the CAA in this refined form.

Vertical motion at radius $R$ is controlled by the potential $\Psi_z(z) =
\Phi(R, z)-\Phi(R,0)$, so the vertical energy is
 \begin{equation}
E_z=\half v_z^2+\Psi_z(z)
\end{equation}
and the vertical action is
\begin{equation}
J_z = \frac{2}{\pi}\int_{0}^{z_{\rm max}}\mathrm{d}z\,v_z,
\end{equation}
 where $z_{\rm max}$ is the height above the plane where $v_z$ vanishes.  On
a grid in $(R,E_z)$ we use these formulae to tabulate $J_z$, and by
interpolation in this table can recover $E_z(J_z,R)$. The assumption that
$J_z$ is constant along the orbit, then yields the variation of $E_z$ with
$R$. Conservation of radial plus vertical energy implies that the radial
motion is governed by the effective potential
\begin{equation}
\Psi_R(R) =  \Phi(R,0) + \frac{J_\phi^2}{2R^2}+E_z(J_z,R)-E_z(J_z,\Rc),
\end{equation}
where $\Rc$ is the guiding-centre radius.
Using this effective potential, we estimate the radial action as
\begin{equation}
J_r = \frac{1}{\pi}\int_{\Rp}^{\Ra}\mathrm{d}R\,v_R,
\end{equation}
where $\Rp$ and $\Ra$ are the radii where the radial velocity, $v_R$,
vanishes. For our standard setup we use a linearly-spaced grid of $100$ points in $R_0$ and a linearly-spaced grid of $100$ points in $E_z^2$.

\subsection{Spheroidal adiabatic approximation (SAA)}\label{Method::SAA}

The CAA assumes that stars oscillate parallel
to the $z$ axis with the consequence that orbits are bounded by cylinders
$R=\hbox{constant}$. Orbits that move significant distances from the Galactic
plane are much more nearly bounded by ellipses than straight vertical lines
in the $(R,z)$ plane \citep[e.g.][Figure 3.27]{BinneyTremaine}. Consequently,
we now make the assumption that adiabatically invariant oscillations occur
along the spheroidal surfaces of prolate spheroidal coordinates. We call this
method the spheroidal adiabatic approximation (SAA).

For a general axisymmetric potential, $\Phi(R,z)$ the Hamiltonian in prolate
spheroidal coordinates is [cf eqn~\ref{Eq::Hamiltonian}]
\begin{equation}
H = \half\Big(\frac{p_\lambda^2}{P_\lambda^2}+\frac{p_\nu^2}{P_\nu^2}
+\frac{J_\phi^2}{R^2(\lambda,\nu)}\Big)+\Phi(\lambda,\nu),
\end{equation}
where
\begin{equation}
\begin{split}
P_\lambda^2 &=\frac{\lambda-\nu}{(\lambda-a^2)(\lambda-c^2)}\cr
P_\nu^2 &=\frac{\nu-\lambda}{(\nu-a^2)(\nu-c^2)}.
\end{split}
\end{equation}
Now we assume the `vertical' motion follows an ellipse of constant
$\lambda$ such that the $\nu$ coordinate is determined by the
potential
\begin{equation}
\Psi_\nu(\nu) = \frac{J_\phi^2}{2R^2(\lambda,\nu)}
-\frac{J_\phi^2}{2R^2(\lambda,c^2)}+\Phi(\lambda,\nu)-\Phi(\lambda,c^2).
\end{equation}
The energy of the $\nu$ oscillations is
\begin{equation}
E_\nu = {p_\nu^2\over2P_\nu^2} + \Psi_\nu(\nu)
\end{equation}
and the vertical action is
\begin{equation}\label{eq:SAAJz}
J_z = \frac{2}{\pi}\int_{c^2}^{\nu_+}\mathrm{d}\nu\,p_\nu
= \frac{2}{\pi}\int_{c^2}^{\nu_+}\mathrm{d}\nu\sqrt{2P_\nu^2(\lambda,\nu)}\sqrt{E_\nu-\Psi_\nu(\nu)},
\end{equation}
where $\nu_+$ is the root of $E_\nu=\Psi_\nu(\lambda,\nu)$.  Now given a 6D
phase-space point $(\bs{x},\bs{v})$ we  find the best prolate spheroidal
coordinate system using equation~\eqref{DeltaGuess}, evaluate $E_\nu$ at this point and then obtain $J_z$
by evaluating the integral of equation \eqref{eq:SAAJz} along the curve of constant
$\lambda$ that passes through the given phase-space point.

$J_r$ is determined from
\begin{equation}
J_r =  \frac{1}{\pi}\int_{\lambda_-}^{\lambda_+}\mathrm{d}\lambda\,p_\lambda
= \frac{1}{\pi}\int_{\lambda_-}^{\lambda_+}\mathrm{d}\lambda\sqrt{2P_\lambda^2(\lambda,c^2)}\sqrt{E_{\rm{tot}}-\Psi_\lambda(\lambda)}
\end{equation}
where $\lambda_+$ and $\lambda_-$ are the roots of
$E_{\rm{tot}}=\Psi_\lambda(\lambda)$ and $\Psi_\lambda(\lambda)$ is an
effective radial potential. The latter is defined by requiring that the sum of the
radial and vertical energies is conserved as $E_\nu$ varies along the orbit.
On a grid in $(R,J_\phi,E_\nu)$, where $R$ is the radius at which the curve
of constant $\lambda$ cuts the plane, we tabulate $J_z$ from
eqn~(\ref{eq:SAAJz}). Then by interpolation we can recover $E_\nu$ from given
values of
$(R,J_\phi,J_z)$. Clearly $J_\phi$ and $J_z$ are constant along an
orbit, so for an individual orbit   $E_\nu$ is a function of $\lambda(R)$ alone.
Hence the radial effective potential
\begin{equation}
\Psi_\lambda(\lambda) = \Phi(\lambda,c^2) + \frac{J_\phi^2}{2R^2(\lambda,c^2)}+E_\nu(R=\sqrt{\lambda-a^2},J_\phi,J_z).
\end{equation}
 is well defined. It is straightforward to show that with this definition of
$\Psi_\lambda$, the orbit's energy is
\begin{equation}
E_{\rm tot} =  {p_\lambda^2\over2P_\lambda^2} + \Psi_\lambda(\lambda).
\end{equation}

The grid on which $J_z$ is evaluated is defined in terms of $R$ rather than
$\lambda$ so that we can use different choices of $\Delta$ for each orbit.

\begin{figure}
$$\includegraphics{{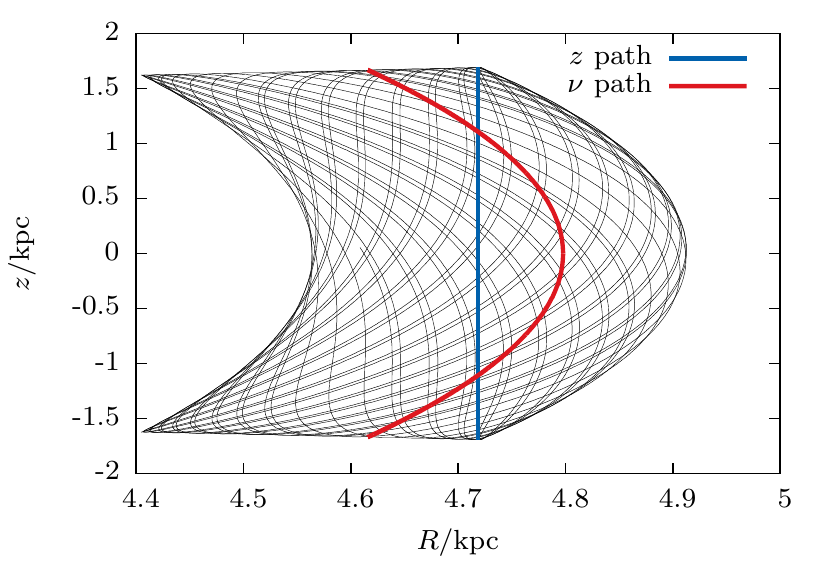}}$$
\caption[Difference between integration paths for the CAA and
SAA]{Illustration of the difference between the CAA and SAA. An example
orbit is shown by the black line. The blue line shows the line along which the CAA vertical-action integration is performed. The red line shows the line of constant $\lambda$ along which the SAA vertical-action integration is performed. The points at which they intersect gives the $(R,\pm|z|)$ coordinates of the input phase-space point.}
\label{SinglePoint}
\end{figure}

An illustration of how this method differs from the CAA is shown in
Figure~\ref{SinglePoint}. We show an example thick disc orbit integrated in a realistic Galactic potential in the meridional plane. The blue line shows the line along which we
integrate to determine $J_z$ with the CAA, while the red line shows the line
of constant $\lambda$ along which we integrate with the SAA. The points at
which they intersect give the $(R,\pm|z|)$ coordinates of the input
phase-space point.  The SAA clearly better captures the shape of the
boundaries of the orbit.

As we will see in Section~\ref{Sect::MethodComparison}, the SAA is
significantly more accurate than the CAA. However, on account of $E_z$ now
being a function of three variables $(R,J_\phi,J_z)$, the SAA takes slightly
longer to use than the CAA. Fortunately, the tabulation of $E_\nu$ requires a very
small number of $J_\phi$ values (here we use ten) over the required $J_\phi$
range. Note that the CAA is contained within the SAA in that the surfaces of
constant $\lambda$ and $\nu$ tend to those of constant $R$ and $z$ in the
limit of very large inter-focal distance. In this limit, the vertical action
becomes independent of $J_\phi$ and the SAA tends to the CAA.

For our standard setup we use a linearly-spaced grid of $100$ points in $\lambda$, a linearly-spaced grid of $100$ points in $E_\nu^2$ and a $10$-point linearly-spaced grid in $J_\phi$.

\subsection{St\"ackel Fudge}\label{Method::SF}

The essence of the adiabatic approximations is the definition of
one-dimensional effective potentials that enable one to obtain $J_z$ and
$J_r$ from quadratures.  \cite{Binney2012} presented an alternative way of
defining one-dimensional potentials, also using prolate spheroidal
coordinates. \cite{SandersBinney2015} extended this idea  to the triaxial
case.

\subsubsection{Axisymmetric case}\label{Method::SF_Axi}

For a general oblate axisymmetric potential,
$\Phi$, we define
\begin{equation}
\begin{split}
\chi_\lambda(\lambda,\nu) &\equiv -(\lambda-\nu)\Phi,\\
\chi_\nu(\lambda,\nu) &\equiv (\lambda-\nu)\Phi.
\end{split}
\end{equation}
If $\Phi$ were a St\"ackel potential, these quantities would be given by
\begin{equation}
\begin{split}
\chi_\lambda(\lambda,\nu) &= F(\lambda)-F(\nu),\\
\chi_\nu(\lambda,\nu) &= F(\nu)-F(\lambda).
\end{split}
\end{equation}
Therefore, for a general potential we can write,
\begin{equation}
F(\tau) \approx \chi_\tau(\lambda,\nu)+D_\tau,
\end{equation}
where $D_\tau$ are constants provided we evaluate $\chi_\lambda$ at constant
$\nu$ and vice versa. We can write the equation for $p_\tau(\tau)$
(eqn~\ref{Eq::EqnOfMotionAxi}) as
\begin{equation}
2(\tau-a^2)(\tau-c^2)p_\tau^2 = (\tau-c^2)E-\Big(\frac{\tau-c^2}{\tau-a^2}\Big)\frac{J_\phi^2}{2}-B_\tau+\chi_\tau(\lambda,\nu),
\label{SFEqnOfMotion}
\end{equation}
where we have defined the integrals of motion $B_\tau = I_3-D_\tau$. Given an
initial phase-space point, we use equation~\eqref{DeltaGuess} to find a
suitable coordinate system, calculate $\lambda,\nu,p_\lambda$ and $p_\nu$,
and use equation~\eqref{SFEqnOfMotion} to find the integrals $B_\tau$. The
momentum $p_\tau$ from equation~\eqref{SFEqnOfMotion} is then integrated over
an oscillation in $\tau$ to find each action as in
equation~\eqref{Eq::action}. We note that for the $\lambda$ integral we keep
$\nu$ fixed at the input value, and vice versa.

\cite{Binney2014_ISO} used an alternative to equation \eqref{DeltaGuess} as a
method of estimating $\Delta$. At each energy and $J_\phi$ a shell orbit $J_r=0$
is computed and an ellipse is fitted to the orbit in the meridional plane.
The location of the focus of the ellipse gives an estimate of $\Delta$. When
$\Delta$ is determined from equation \eqref{DeltaGuess} we shall refer to the
method as the St\"ackel Fudge v1, and when $\Delta$ is determined from shell
orbits as just described, we will call the method St\"ackel Fudge v2.
We test both methods below.

\subsubsection{Triaxial case}\label{Method::SF_Triax}
The above procedure is simply generalized to triaxial potentials. Given a general triaxial potential, we define the quantities
\begin{equation}
\begin{split}
\chi_\lambda(\lambda,\mu,\nu) &\equiv (\lambda-\mu)(\nu-\lambda)\Phi(\lambda,\mu,\nu),\\
\chi_\mu(\lambda,\mu,\nu) &\equiv (\mu-\nu)(\lambda-\mu)\Phi(\lambda,\mu,\nu),\\
\chi_\nu(\lambda,\mu,\nu) &\equiv (\nu-\lambda)(\mu-\nu)\Phi(\lambda,\mu,\nu).
\end{split}
\end{equation}
If $\Phi$ were a St\"ackel potential, these quantities would be given by, for instance,
\begin{equation}
\chi_\lambda(\lambda,\mu,\nu) = F(\lambda)-\lambda\frac{F(\mu)-F(\nu)}{\mu-\nu}+\frac{\nu F(\mu)-\mu F(\nu)}{\mu-\nu}.
\end{equation}
Therefore, for a general potential, we can write
\begin{equation}
F(\tau) \approx \chi_\tau(\lambda,\mu,\nu)+C_\tau\tau+D_\tau,
\end{equation}
where $C_\tau$ and $D_\tau$ are constants provided we always evaluate
$\chi_\tau$ with two of the ellipsoidal coordinates fixed. For instance, we
always evaluate $\chi_\lambda$ at fixed $\mu$ and $\nu$.

When we substitute these expressions into equation~\eqref{Eq::EqnOfMotion}, we find
\begin{equation}
2(\tau-a^2)(\tau-b^2)(\tau-c^2)p_\tau^2=\tau^2 E -\tau A_\tau+B_\tau +\chi_\tau(\lambda,\mu,\nu).
\label{Eq::EqnOfMotion_JK}
\end{equation}
For each coordinate $\tau$, there are two new integrals of motion given by
$A_\tau=a-C_\tau$ and $B_\tau=b+D_\tau$. Using a single 6D coordinate and a
choice of coordinate system gives us a single constraint on a combination of
$A_\tau$ and $B_\tau$. Due to the separability of St\"ackel potentials the
derivative of the Hamiltonian with respect to the ellipsoidal coordinates
will be zero for a true St\"ackel potential. Setting it equal to zero for a
general potential gives a further constraint on $A_\tau$ and $B_\tau$ allowing
us to solve for these integrals given only a single $(\bs{x},\bs{v})$
coordinate. Then we have an approximate expression for $p_\tau(\tau)$ that
can be integrated to estimate the actions.

\subsection{St\"ackel fitting}\label{Method::FIT}

Whereas the St\"ackel Fudge uses the real potential as if it were a St\"ackel
potential, \cite{Sanders2012a} explicitly fitted a St\"ackel potential to the
real potential. This method uses the procedure from
\cite{DejonghedeZeeuw1988} to find the locally best-fitting St\"ackel
potential. One minimises the difference between the auxiliary function
$(\lambda-\nu)\Phi$ formed from the true potential and the St\"ackel
auxiliary function $F(\nu)-F(\lambda)$. The best-fitting $F$ can be computed
by an integral over a region of the prolate spheroidal coordinate system
that is defined on an orbit-by-orbit basis. We first integrate the orbit
for several time-steps to form an average estimate of $\Delta^2$ via
equation~\eqref{DeltaGuess}. We further integrate the orbit to determine the
orbit boundaries in $\lambda$ and $\nu$. With the boundaries found we compute
the best-fitting $F$ on a grid (we use $40$ grid points) and hence the
best-fitting St\"ackel potential in which the actions can be computed as
detailed in Section~\ref{StackelPot}.

The fitting procedure of \cite{DejonghedeZeeuw1988} uses weight functions
which allow some flexibility in the fitting of the potential. We use weight
functions $\Lambda(\lambda)\propto (\lambda-\Delta^2)^{-4}$ and
$N(\nu)\propto (\nu-c^2)^{0.5}$.

\subsection{Angle and frequency estimation}\label{Sect::AngleFreq}

The angle
coordinates are related to the generating function, $S$, by
\begin{equation}
\btheta = \frac{\partial S}{\partial \bs{J}}.
\end{equation}
In an axisymmetric  St\"ackel  potential this generating function has the form
\begin{equation}
S(\lambda,\phi,\nu,J_r,J_\phi,J_z) = \int^\lambda\mathrm{d}\lambda\,p_\lambda
+\int^\phi\mathrm{d}\phi\,p_\phi+\int^\nu\mathrm{d}\nu\,p_\nu.
\end{equation}
 The angle coordinates can be computed by writing
\begin{equation}
\theta_i = \sum_k\frac{\partial S}{\partial I_k}\frac{\partial I_k}{\partial J_i}.
\end{equation}
 The first term is computed by differentiating the
momenta $p_i$ with respect to the classical integrals $I_k$ and
integrating along the coordinate path. The second term is computed by
inverting the matrix $\partial J_i/\partial I_k$ which again is found
by differentiating the momenta in the action integrals and integrating over
the full orbital path  \citep[see for example the appendix of][]{Sanders2012a}.
This approach can be extended to compute the angles using the presented
approximation schemes. In the adiabatic approximations the three integrals
are taken to be $(E,J_\phi,J_z)$. When computing the derivatives of the
generating function and $J_r$ with respect to $J_z$ we require the
derivatives $\partial J_z/\partial E_z|_R$ and $\partial J_z/\partial
E_\nu|_{\lambda,J_\phi}$ which we choose to compute from the grids (note these quantities can be computed by integration but as they are required for each integrand call this would be computationally expensive). In the
St\"ackel Fudge the classical integrals are $(E,J_\phi,B_\tau)$ and finally
in the St\"ackel fitting procedure the classical integrals are $(E,J_\phi,I_3)$.

The frequencies are given by
$\partial H/\partial \bs{J}$, so these are just single components of the
inverse of the matrix $\partial J_i/\partial I_k$.

There is a significant difference between how the actions and the frequencies
are evaluated. Each action is an integral of the momenta, while a frequency
is the integral of the inverse of the momentum. Therefore, the error in the
action is dominated by the regions of high momentum in the middle of the
range of integration while the error in the frequency is dominated by errors
in the integrand near the end points of the range of integration, where the
integrand diverges. For this reason, the frequencies are much more sensitive
to the location of the orbit's turning points while the actions are more
sensitive to how closely the potential is modelled within the body of the orbit.

\section{Method comparisons}\label{Sect::MethodComparison}

In this section we critically compare the methods presented
above. We focus on two quantities for each method: the accuracy of
the action computation (i.e. how constant are the actions around an orbit)
and how much computing time is required. The latter quantity is
slightly subjective as it depends on  computational
details. However, we will see that the time differences between the methods
are orders of magnitude, so we do not expect programming upgrades to disturb
the rank-ordering of the methods.

Since there are more competing methods for the
axisymmetric than for the triaxial case,
we limit our comparison to the axisymmetric versions of the
algorithms -- in the triaxial case the choice currently is between
the O2GF method of Section~\ref{Method::Genfunc} if one
requires a few accurate actions and the St\"ackel Fudge method of
Section~\ref{Method::SF_Triax} if one requires many actions with a lower
level of accuracy. The  axisymmetric methods we compare are
 \begin{enumerate}
\item Cylindrical adiabatic approximation (CAA), Section~\ref{Method::PAA}
\item Spheroidal adiabatic approximation (SAA), Section~\ref{Method::SAA}
\item St\"ackel Fudge with $\Delta$ chosen using equation~\eqref{DeltaGuess} (Fudge v1), Section~\ref{Method::SF}
\item St\"ackel Fudge with $\Delta$ chosen from shell orbits (Fudge v2), Section~\ref{Method::SF}
\item Locally fitting St\"ackel potentials (Fit), Section~\ref{Method::FIT}
\item Iterative torus construction (ItTC), Section~\ref{Method::ItTorus}
\item Computing the generating function  from an orbit (O2GF), Section~\ref{Method::Genfunc}
\item Averaging toy actions over toy angles (AvGF), Section~\ref{Method::AvGenfunc}
\end{enumerate}
 We begin by inspecting numerically estimated actions at a series of times
along single orbits (Section~\ref{SingleOrbs}) and we go on to investigate
the action, angle and frequency variances for a broader range of orbits
(Section~\ref{ManyOrbs}).

\subsection{Four representative orbits}\label{SingleOrbs}

\begin{figure*}
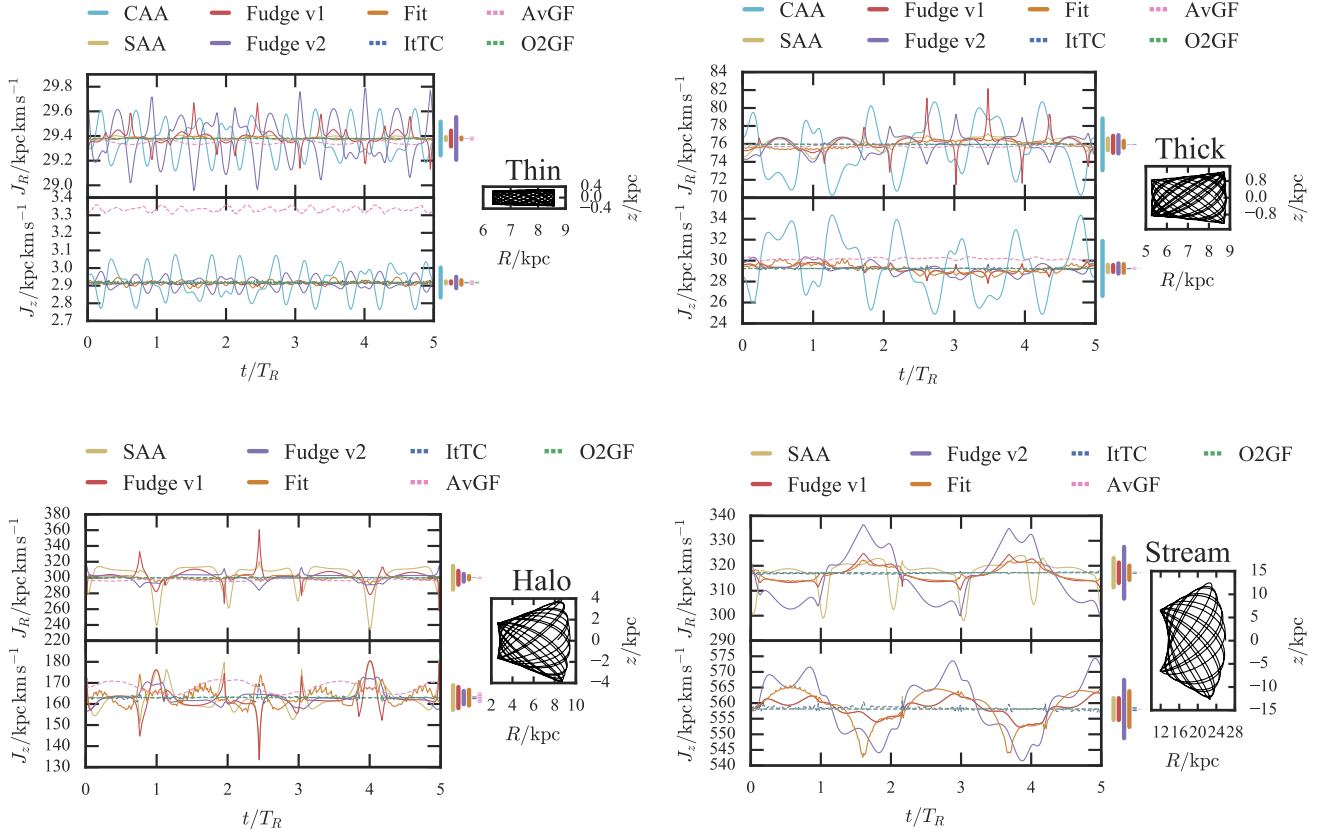

\centering
\begin{minipage}{.5\textwidth}
\centering
$$\includegraphics[width=\textwidth]{{{plots/thin.action}}}$$
\end{minipage}%
\begin{minipage}{0.5\textwidth}
\centering
$$\includegraphics[width=\textwidth]{{{plots/thick.action}}}$$
\end{minipage}
\begin{minipage}{.5\textwidth}
\centering
$$\includegraphics[width=\columnwidth]{{{plots/halo.action}}}$$
\end{minipage}%
\begin{minipage}{0.5\textwidth}
\centering
$$\includegraphics[width=\columnwidth]{{{plots/stream.action}}}$$
\end{minipage}
\caption{Comparison of action estimation methods for single orbits.
In each set of panels the orbit's trace in the
meridional plane $(R,z)$ is shown on the right.  The set
of panels at top left is for a typical \emph{thin} disc orbit, the set at top
right is for a typical \emph{thick} disc orbit, that at bottom left is for a
typical \emph{halo} orbit, and the bottom right set is for a typical orbit of
the progenitor of a tidal \emph{stream}.  Each set of panels shows on the
left the radial action, $J_r$, (top panel) and the vertical action, $J_z$,
(lower panel) as a function of time in units of the radial period. Values
from \emph{non-convergent} methods are shown with solid lines whilst values
from \emph{convergent} methods are shown with dashed lines. Cyan solid line
is for the CAA, mustard-yellow solid is for the SAA, red solid line is for the
St\"ackel Fudge with variable $\Delta$ estimation (v1), the purple solid is for
the St\"ackel Fudge with $\Delta$ estimated from shell orbits (v2), the solid
orange line is for the St\"ackel fitting method, the blue dashed line
is for the ItTC method, the green dashed line is for the O2GF method,
and the pink dashed line is for the AvGF method. To the right of each
sub-panel the vertical lines show the $\pm$ standard deviation spread in the
action estimates from each method. All orbits are computed in
the multi-component potential from \protect\cite{Piffl2014}.}
\label{Fig::SingleOrbs}
\end{figure*}

\begin{table*}
\caption{Errors in the action for four example orbits: on the first line we give the
radial and vertical actions for the four orbits in units of $\!\kpc\kms$. Below
the double horizontal separators we give the RMS deviations of the
radial and vertical action estimates relative to the mean of the O2GF estimates for the eight methods. The methods above
the horizontal separator are the non-convergent methods whilst those below
are the convergent methods.} \begin{tabular}{llrrrrrrrr}
&&\multicolumn{2}{c}{Thin}&\multicolumn{2}{c}{Thick}&\multicolumn{2}{c}{Halo}&\multicolumn{2}{c}{Stream}\\
&&$J_R$&$J_z$&$J_R$&$J_z$&$J_R$&$J_z$&$J_R$&$J_z$\\
\hline
 &Actions & 29.38 & 2.92 & 75.96 & 29.25 & 299.79 &  163.02 & 317.20 &   558.09 \\
\hline
\hline
 \textbf{Method}&CAA      & 0.1   & 0.08  & 3   & 3   & 40 & 50 & 100 & 200 \\
 &SAA      & 0.01  & 0.007 & 0.7 & 0.4 & 10 & 6     & 6     & 4     \\
 &Fudge v1 & 0.07  & 0.007 & 1   & 0.3 & 9     & 5     & 4     & 3     \\
 &Fudge v2 & 0.2   & 0.03  & 1   & 0.5 & 5     & 3     & 10 & 9     \\
 &Fit      & 0.007 & 0.01  & 0.4 & 0.4 & 2     & 4     & 3     & 6     \\
\hline
 &ItTC   & 0.002  & 0.008 & 0.03  & 0.06  & 0.3  & 0.8  & 0.4  & 0.6  \\
 &AvGF & 0.03   & 0.4   & 0.2   & 2     & 3    & 6    & 0.2  & 0.3  \\
 &O2GF   & 0.0003 & 0.005 & 0.002 & 0.007 & 0.03 & 0.03 & 0.01 & 0.01 \\
\hline
\end{tabular}
 \label{Table}
\end{table*}

We
analyse results for four representative Galactic orbits: a typical
\emph{thin} disc orbit; a typical \emph{thick} disc orbit; a typical
\emph{halo} orbit; and a typical orbit for the progenitor of a tidal
\emph{stream}. The initial conditions of each orbit are:
\begin{enumerate}
\item Thin,
$$\bs{x} = (8.29,0.1,0.1)\kpc, \bs{v}=(30.22,211.1,19.22)\kms,$$
\item Thick, $$\bs{x} = (8.29,0.1,0.1)\kpc, \bs{v}=(50.22,187.1,54.22)\kms,$$
\item Halo, $$\bs{x} = (8.29,0.1,0.1)\kpc, \bs{v}=(100.22,109.1,101.22)\kms,$$
\item Stream, $$\bs{x} = (26.,0.1,0.1)\kpc, \bs{v}=(0.1,141.8,83.1)\kms.$$
\end{enumerate}
The thin, thick and halo orbits were chosen arbitrarily but the stream orbit
is selected as a likely orbit for the progenitor of the GD-1 stream
\citep{Koposov2010,SandersBinney2013b}. Each orbit was integrated for $10$
orbital periods of a circular orbit of the same energy and $1000$ time
samples were recorded.  The orbits were integrated in the multi-component
Galactic potential of \cite{Piffl2014}. This potential was fitted to the
kinematics of RAVE stars, the vertical density profile of the solar cylinder
measured by the SDSS survey \citep{Juric2008}, and the constraints on the
circular speed as a function of Galactocentric radius assembled by
\cite{McMillan2011}.  The generating mass distribution consists of three
exponential discs (thin, thick and gas), a central bulge and an NFW dark
halo. The actions were computed using the eight different methods for each
time sample.  These estimates are plotted in Fig.~\ref{Fig::SingleOrbs} and
their standard deviations are given in Table~\ref{Table}.

The ItTC and O2GF methods produce RMS deviations
that are small enough for all purposes for all the inspected orbits, so we
will not comment on them further in this section.

\subsubsection{Thin}

From the plot of the orbit in the meridional plane, we can clearly see that
the orbit is nearly separable in cylindrical polar coordinates. However, the
St\"ackel Fudge and SAA  methods give better results than the CAA. Of the
non-convergent methods, St\"ackel fitting gives the most accurate results.

For this very low eccentricity orbit, locally choosing $\Delta$ around the
orbit (Fudge v1) gives better results than using shell orbits to choose the
parameter $\Delta$ of the ellipsoidal coordinate system (Fudge v2), but this
ordering probably reflects the grid used to compute $\Delta$ for the closed
orbits.

The AvGF method gives slightly biased estimates of both actions. One
reason for this is that the toy actions are always positive, so for small
actions, averaging produces a positively biased action.  This might be improved
by using a better toy potential but we do not explore this option here.

\subsubsection{Thick}

The left and right boundaries of the thick disc orbit show significant
curvature, so the assumption of separability in cylindrical polar coordinates
is not a good one. For this reason the CAA is inferior to
the other methods by a factor of a few. All four of the methods based on
St\"ackel potentials give comparable accuracy for $J_z$ but explicit St\"ackel
fitting gives a result that is twice as  accurate as the other methods for
$J_r$. As for the thin-disc orbit, the AvGF method gives slightly
biased estimates of both $J_r$ and $J_z$, with the error in $J_z\sim
1\kpc\kms$.

\subsubsection{Halo}

For the halo and stream orbits we have not plotted results from the CAA
because the latter is not competitive at these higher eccentricities.  The
hierarchy for the St\"ackel-type methods is SAA, Fudge v1, Fudge v2 and
St\"ackel-fitting with the St\"ackel-fitting method producing a value of
$J_r$ that is a factor of five more accurate than that from the SAA.  All
three methods produce equally accurate results for $J_z$. The AvGF
method again produces a small systematic shift in $J_z$.

Since on this orbit the $z$ oscillations do not occur faster than the
$R$ oscillations, there is no reason why $J_z$ should be adiabatically
conserved as the SAA assumes. Hence it is remarkable that the SAA performs
quite well on this orbit.

\subsubsection{Stream}

 The stream-like orbit ranks the St\"ackel-type methods in the same order as
the halo orbit. The St\"ackel Fudge v1 and the SAA produce very similar
results for both $J_r$ and $J_z$. St\"ackel-fitting produces more accurate
results for $J_r$ but a less accurate value for $J_z$ and the St\"ackel
Fudge v2 yields slightly larger errors in both $J_r$ and $J_z$. Evidently,
estimating $\Delta$ from shell orbits is not optimal for this more
radially-extended orbit. There are no obvious systematic shifts in the
actions from the AvGF method.

\begin{figure}
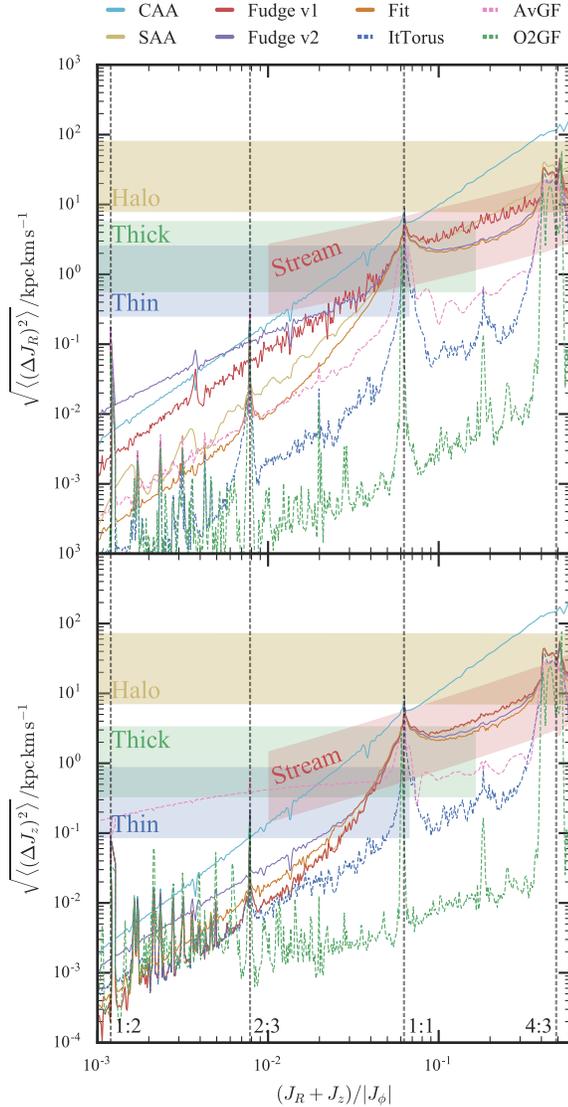

$$\includegraphics[width=\columnwidth]{{{plots/many_tori_output.dat.acc}}}$$
\caption{Comparison of the accuracy of the action-estimation methods for
multiple orbits: each panel shows the root-mean-squared deviations of the
estimates from the mean computed using the O2GF method. The upper panel shows
the radial action $J_r$ and the lower panel shows the vertical action $J_z$.
The \emph{non-convergent} methods are shown with solid lines whilst the
\emph{convergent} methods are shown with dashed lines. The cyan solid line is
for the CAA, the mustard-yellow solid line is for the SAA, the red solid line is for
the St\"ackel Fudge with variable $\Delta$ estimation, the purple solid line
is for the St\"ackel Fudge with $\Delta$ estimation from shell orbits, the
solid orange line is for the St\"ackel fitting method, the blue
dashed line is for the ItTC method, the green dashed line is for the
O2GF method and the pink dashed line is for the AvGF method.  The coloured
bands show the region in which the relative error in the action-based
distribution functions for the thin (blue), thick (green) and halo (yellow)
components are between $1$ and $10\percent$. The upper and lower boundaries
of the red band correspond to the action-space widths of a stream shed
from a progenitor that has
velocity dispersions of $5$ and $0.5\kms$. The chosen horizontal limits of
these bands are $\log f/f_{\rm max}=-5$ for the thin and thick components and
$(J_r+J_z)/|J_\phi|=0.01$ for the stream component. The vertical grey dashed
lines correspond to resonances $x:y$ when $\Omega_R/\Omega_z=x/y$.}
\label{MultiOrbits}
\end{figure}

\begin{figure}
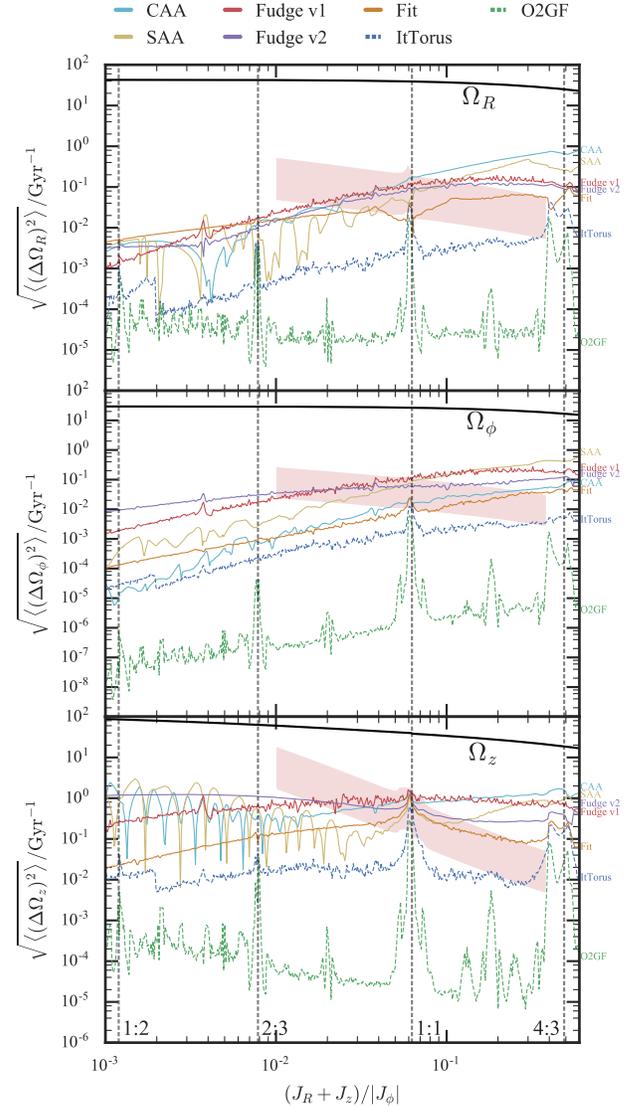

$$\includegraphics[width=\columnwidth]{{{plots/many_tori_output.dat.freq}}}$$
\caption{Comparison of the accuracy of frequencies estimated using different
methods: the top panel shows the radial frequency, middle panel the azimuthal
and bottom panel the vertical frequency. Each panel shows the root-mean-squared deviations of the estimates from the mean computed using the O2GF method. The colour-coding and range of
orbits explored are as in Fig.~\protect\ref{MultiOrbits}. The solid black lines show the frequencies of the orbits computed using the O2GF method. The upper and lower
boundaries of the red band correspond to the frequency widths of streams
with velocity dispersions of $5$ and $0.5\kms$. The vertical grey dashed
lines correspond to resonances $x:y$ when $\Omega_R/\Omega_z=x/y$.}
\label{MultiOrbits_Freq}
\end{figure}

\begin{figure}
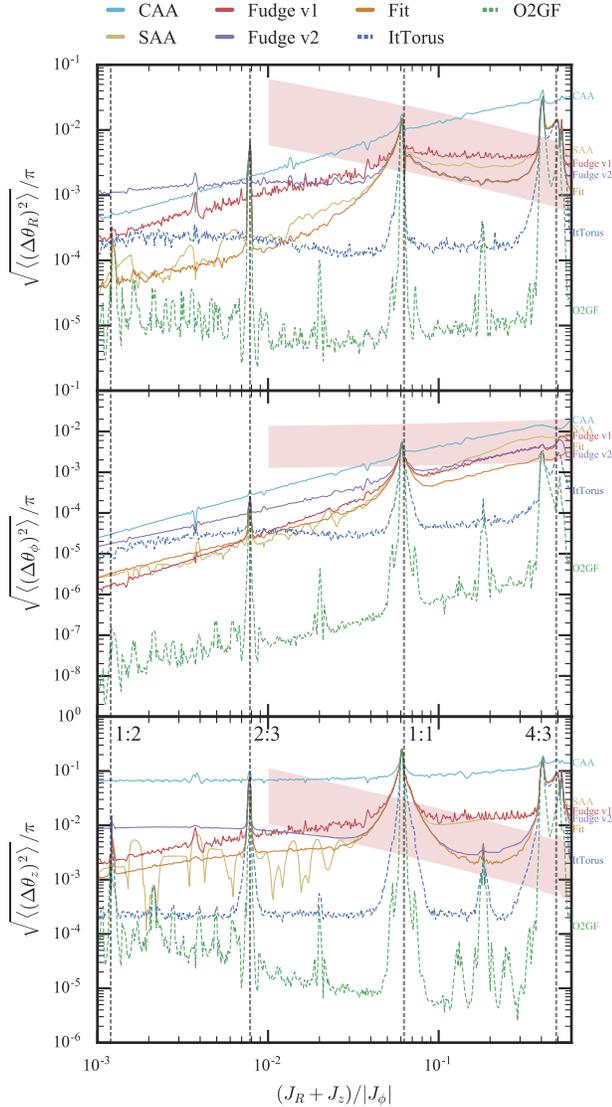

$$\includegraphics[width=\columnwidth]{{{plots/many_tori_output.dat.ang}}}$$
\caption{Comparison of the accuracy of angle estimation using different
methods for multiple orbits: the top panel shows the radial angle, middle
panel the azimuthal and bottom the vertical. The colour-coding and the range
of orbits explored are as in \protect\ref{MultiOrbits}. Each panel shows the root-mean-squared deviations of the estimates from the angles computed using the O2GF method. The O2GF line shows the root-mean-squared deviations from the angles computed using $\theta_i(0)+\langle\Omega_i\rangle t$. The upper and lower
boundaries of the red band correspond to the angle widths of streams with
scale radii of $100$ and $10\pc$. The vertical grey dashed lines correspond
to resonances $x:y$ when $\Omega_R/\Omega_z=x/y$.} \label{MultiOrbits_Ang}
\end{figure}

\subsection{Multiple orbits}\label{ManyOrbs}

We now analyse results for a larger sample of orbits. All orbits have
$J_\phi=R_0V_c(R_0)$, where $R_0$ is the solar radius and $V_c$ is the
circular speed. At a series of steadily increasing energies we set
$v_z=0.8v_R$ and launch the orbit from the solar position. We integrate
orbits for ten periods of the circular orbit with the same energy and compute
the actions, angles and frequencies at $40$ points along the orbit.  The most
extreme orbit considered has pericentre $\sim5\kpc$, apocentre $\sim28\kpc$
and maximum Galactic height $\sim15\kpc$.

In Figure~\ref{MultiOrbits} we show as a function of $(J_r+J_z)/|J_\phi|$,
which is a measure of eccentricity, the RMS of the differences between the
actions from a particular method and the mean of the values obtained from the
O2GF method. The use of the RMS does not give the full story as inspection of Figure~\ref{SingleOrbs} shows that the action fluctuations can be quite peaky, particularly around the turning points of the orbit. However, it serves as a useful summary statistic. Since the green dashed curve of the O2GF method runs along the
bottom of both the upper panel for RMS in $J_r$ and the lower panel for RMS
in $J_z$, the O2GF method emerges as the most accurate method regardless of
orbital eccentricity. It must, however, be acknowledged that our playing
field is not entirely level in that RMS fluctuations are computed relative to
the mean of the O2GF results, so for O2GF to show no RMS it merely needs to
extract the same actions from two, strongly overlapping segments of orbit.
In particular, its RMS would be zero even if it consistently returned
erroneous actions.

The next most accurate method is the ItTC method. As regards $J_z$ it
is followed by the St\"ackel Fudge v1 and SAA methods at small eccentricities and by the AvGF method at larger eccentricities. However, consideration of $J_r$ yields a
different ordering with both the St\"ackel fitting and AvGF methods
proving more accurate than the St\"ackel Fudge v1 at all eccentricities. The SAA and St\"ackel Fudge methods calculate the vertical action in very similar ways so at all eccentricities St\"ackel Fudge v1 method has near identical errors to the SAA method. However, inspection of Figure~\ref{SingleOrbs} shows that the individual action estimates around an orbit are not identical for the two methods. At high eccentricities the St\"ackel fitting method produces the best results of the non-convergent methods followed by Fudge v2 and then the SAA and Fudge v1. At the largest eccentricities the CAA produces errors that are an order of magnitude larger than the St\"ackel-based non-convergent methods.

In Fig.~\ref{MultiOrbits} vertical lines mark eccentricities at which
orbits may become trapped by resonances. At these eccentricities there are
sharp peaks in the errors for all methods. The peaks are particularly sharp
for the O2GF method. For eccentricities smaller than that at which the 2:3
resonance occurs, the errors in $J_z$ are similar for the O2GF, ItTC,
and St\"ackel Fudge methods. Even at these low eccentricities the errors in
$J_r$ from the St\"ackel Fudge v1 are significantly larger than those from
the two convergent methods.

In Fig.~\ref{MultiOrbits_Freq} we show the RMS deviations of the frequencies
from the mean of the frequencies returned by the O2GF method, and in
Figure~\ref{MultiOrbits_Ang} we show the RMS deviation of the angles from
those computed with the O2GF method. Again the O2GF
method shows clear superiority at all eccentricities, followed by the
ItTC method. Of the St\"ackel-based non-convergent methods the broad picture is that St\"ackel fitting performs best followed by the St\"ackel Fudge, the SAA and finally the CAA. An exception to this is the azimuthal frequency $\Omega_\phi$ where the SAA performs worst and the CAA is competitive with St\"ackel fitting. Additionally, the SAA angle estimation is very competitive at low eccentricities. The vertical frequency estimates for both the CAA and SAA are peaky due to the interpolation grid that is used. Again we see a striking deterioration in the performance
of the O2GF method at resonances. The deleterious impact of resonances is
particularly marked in the case of the angle variables.

We have also computed RMS deviations for two series of orbits generated in
the same way as those used for Figs.~\ref{MultiOrbits} to
\ref{MultiOrbits_Ang} but starting from the circular orbits at $5\kpc$ and at
$13\kpc$ rather than at $R_0$. The results are qualitatively the same as
those given above.  We further explored the impact on these tests of
replacing the realistic Galactic potential by a flattened
logarithmic potential ($q=0.9$). The errors with the actions are much
smoother in this case and resonances are not evident. The magnitude of
the errors in the actions are very similar to those in the more realistic
potential with all the non-convergent St\"ackel methods producing very similar magnitude errors that are a factor of ten smaller errors than the CAA, and the AvGF and ItTC methods producing errors another factor of ten smaller for the most eccentric orbits. The O2GF method produces significantly smaller errors in the logarithmic potential than in the multicomponent potential ($\sim$ a factor
of ten) as the smoothness of the logarithmic potential means fewer terms are
required in the generating function.

\subsection{Computational time}

\begin{figure}
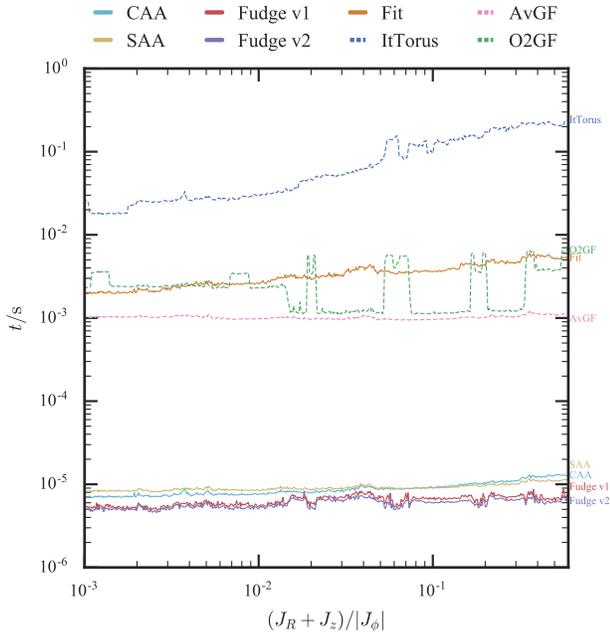

$$\includegraphics[width=\columnwidth]{{{plots/many_tori_output.dat.times}}}$$
\caption{Comparison of the speed of the methods. The colour-coding and range
of orbits explored are as in Fig.~\protect\ref{MultiOrbits}. The time quoted is
the time per estimated action averaged from $50$ estimates per orbit.}
\label{MultiOrbits_Time}
\end{figure}

In Figure~\ref{MultiOrbits_Time} we show for each method the average time for
computing a single action as a function of $(J_r+J_z)/|J_\phi|$. The actions were computed on a single $4\mathrm{GHz}$ Intel i7 processor. For an orbit
of given eccentricity, these times vary by more than three orders of
magnitude, with the St\"ackel Fudges being cheapest ($\sim5\times10^{-6}\mathrm{s}$)
and the ItTC method most expensive. In the bargain basement we have
the St\"ackel Fudges and the adiabatic approximations CAA and SAA.  Given
that the CAA and SAA are two to three times more expensive (because they
involve interpolations in two- and three-dimensional tables), and are never
significantly more accurate (Fig.~\ref{MultiOrbits} and Table~\ref{Table})
than the St\"ackel Fudges, it is clear that anyone shopping for a cheap and
cheerful method of action determination should choose a St\"ackel Fudge.
Versions 1 and 2 of the St\"ackel Fudge have identical cost but version 1 is
the more accurate except for highly eccentric orbits, when its disadvantage
is marginal. Consequently for general low-precision work St\"ackel Fudge v1
is the clear recommendation.

Fig.~\ref{MultiOrbits_Time} shows the mid-price methods to be the AvGF,
St\"ackel Fit and O2GF. Of these, the O2GF method clearly stands out as the
best buy because it is orders of magnitude more accurate than its peers. Its
only drawback is that it is a complex method to code. The dashed green curve
plotting its cost is jagged because for nearly resonant orbits a longer or
more densely sampled orbit integration is required.

The ItTC method is roughly an order of magnitude more expensive than
any other method. Its strength is robustness in the presence of resonant
trapping.  Each torus construction involves inverting two matrices (one for
the $S_{\bs{n}}$ and another for their derivatives) and we have allowed a
maximum of five torus constructions.  In some applications a significant
speed-up should be possible by  constructing tori by interpolating on a grid
of pre-computed tori as described by \cite{BinneyMcMillan2015}.

\section{Discussion}\label{Sect::Discuss}

We have seen that the costs and performance of action finders vary
considerably, with the St\"ackel Fudge v1 delivering lots of medium-quality
actions in the time required for the O2GF method to deliver a single
high-quality result. Consequently, it is important to understand what
accuracy one really requires for a particular application.

An important application is the computation of observable quantities such as
stellar densities and kinematics from a model with a specified DF $f(\vJ)$.
In this case the key criterion is the accuracy with which either $f(\vJ)$ or
its moments can be evaluated.  \cite{BinneyMcMillan2011} compared the
recovery of the velocity distributions and density profiles of model stellar
discs when the actions were obtained from either an early version of the CAA
or torus mapping. They found that within $\sim1\kpc$ of the Galactic plane
the differences were very small, but further from the plane the adiabatic
approximation produced results that are in error by $\sim10\percent$.
Similarly, \cite{BovyRix2013} showed a comparison between an early version of
the CAA and the St\"ackel Fudge. They found that the difference in the
vertical velocity dispersions obtained from the two methods was $\sim1$ per
cent in a model with a very small vertical gradient in velocity dispersion.

\cite{Binney2014_ISO} showed that densities and velocity dispersions computed for
flattened isochrone spheres using the St\"ackel Fudge v2 satisfy the Jeans
equations to good precision. Similarly,
\cite{SandersBinney2014} demonstrated that actions estimated with the
triaxial St\"ackel Fudge are sufficiently accurate for the resulting density
and velocity dispersions of a triaxial Navarro-Frenk-White dark-matter halo
to satisfy the Jeans' equations. These studies demonstrate that although the
error in the actions for individual orbits can be large for some methods, the
resulting moments of the distribution function are well recovered as they are
presumably dominated by those low-action orbits that have good action
estimates.

In Appendix~\ref{Appendix} we give details of four distribution functions
that describe components of our Galaxy. With these distribution functions we
are able to compare the accuracy of the action calculations required to
evaluate the distribution functions to a certain accuracy. In each panel of
Fig.~\ref{MultiOrbits} we show four coloured bands. The three labelled
\emph{thin}, \emph{thick} and \emph{halo} give the accuracy required to
evaluate the corresponding distribution function accurate to between $1$ and
$10\percent$.  In Fig.~\ref{MultiOrbits} the band labelled \emph{stream} shows the
action-space spreads for a stream created from a progenitor with a velocity
dispersion of between $0.5$ and $5\kms$.  The yellow bands in
Figs.~\ref{MultiOrbits_Freq} and \ref{MultiOrbits_Ang} show the width in
frequencies and angles of particles very recently stripped from such a
progenitor. Since the spread in the angles grows linearly in time as the
stream grows, the given width is the smallest one would ever wish to resolve.
Wherever the curves showing a method's accuracy lie below all these coloured
bands, that method provides sufficient accuracy for the job in hand. We see
that the St\"ackel Fudge v1 comes too high for safety only in a small range
of values of $(J_r+J_z)/|J_\phi|$, so even this cheap and cheerful method is
adequate for all but a few thick-disc orbits.

The spread in frequencies and angles we have computed is averaged over many
orbital periods. In reality, when modelling a stream the stream particles are
confined to a much narrower range of angles. The frequency estimation for the
collection of particles may be systematically offset from the truth due to
the phase of the stream but we anticipate the inter-particle frequency
separations will be well resolved. Therefore, we can perhaps get away with
slightly larger errors in the frequency estimation than those suggested by
the \emph{stream} band in Figure~\ref{MultiOrbits_Freq}.

\subsection{Trapping by resonances}

We have seen that results from different methods tend to diverge for orbits
on which $\Omega_r{:}\Omega_z$ approaches a simple ratio such as 1:2 or 2:3.
This finding arises because these orbits have become trapped by a resonance.
The pre-requisite to a clear discussion of resonant trapping is the existence
of a Hamiltonian $H_0$ that is similar to the actual Hamiltonian $H$ but in which
resonant trapping does not occur. That is, for an  orbit $\vJ^{(0)}$ in
$H_0$,
$\Omega_r/\Omega_z=m/n$ for small integers $m$ and $n$, but on neighbouring
orbits $\Omega_r/\Omega_z$ is an irrational number close to $m/n$. Given the
existence of $H_0$ one can then understand resonant trapping by considering
the impact on orbits in $H_0$ of the small perturbation $\Delta\equiv H-H_0$
\citep[e.g.][]{Binney2013}.

Perturbation theory allows one to estimate the extent of the region in action
space within which the perturbation $\Delta$ will cause orbits to become
trapped by the $m{:}n$ resonance. It may happen that this region overlaps
with the region of entrapment by another resonance, associated with different
integers. In this case, $\Delta$ causes orbits in $H_0$ that lie in the
overlap region to become chaotic in the sense that they cease to be
quasi-periodic \citep{Chirikov1979,Binney2013}.

Resonantly trapped quasi-periodic orbits have actions, but we cannot take
two of these actions to be $J_r$ and $J_z$; instead we must use a linear combination
of $J_r$ and $J_z$ complemented by a new action that quantifies the extent of
their libration about the underlying resonant orbit \citep[see section 3.7.2 of][]{BinneyTremaine}. Chaotic orbits do not
admit a full complement of action integrals.

The applications we have in mind for actions assume that negligibly few
orbits are resonantly trapped, so all orbits can be characterised by
$J_r,J_\phi$ and $J_z$.  Consequently, what we ask of a method of action
determination, is that when it is confronted by a trapped orbit, it returns
the actions of one of the orbits in $H_0$ that is trapped into the given
orbit. We say ``one of'' because a given orbit in $H_0$ can be trapped into
many different orbits in $H$, and a given orbit in $H$ can be reached from
many orbits in $H_0$. Hence resonant trapping obliges us to allow a certain
fuzziness in the meaning of actions. If the \df\ returns similar values for
all actions in the range of a given region of entrapment, this fuzziness is
likely to be unimportant. Thus, the criterion for resonant trapping to be
unimportant is closely related to the criterion we have explored for errors
from action finders being unimportant.

The nearby integrable Hamiltonian $H_0$ can be constructed by torus mapping
\citep{KaJJB94:PhysRev,BinneyMcMillan2015},
and we hope shortly to present practical details of how this is best
achieved.

\subsection{Convergence of the O2GF method}

At the start of the previous section we remarked that the two convergent
methods produced results that by some distance out-rank the results from the
non-convergent methods. However, the results we gave were obtained using just
one set of parameters in each convergent method.  Here we explore how the
error in the actions changes as a function of the parameters used in the O2GF
method.

There are three parameters to choose for the O2GF method: the
integration time $N_T$, the number of samples $N_\mathrm{samp}$ and the
number of Fourier coefficients (controlled by the parameter
$N_\mathrm{max}$). The error in the action proves to be
most strongly a function of the number of Fourier coefficients. For the
frequencies and angles the error is also affected by the total integration
time.

\begin{figure}
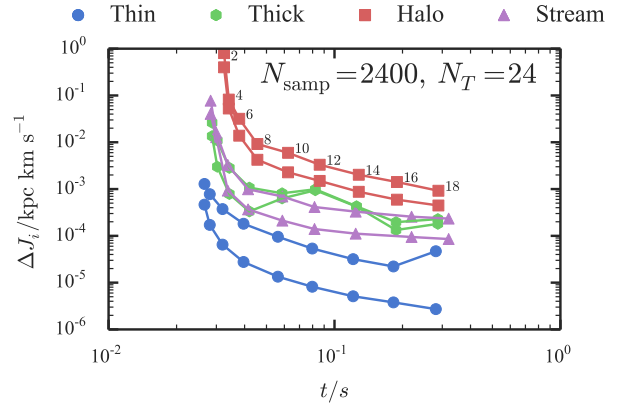

$$\includegraphics[width=\columnwidth]{{{plots/genfunc_converg}}}$$
\caption{Error in the actions as a function of computation time for the
O2GF method using different numbers of Fourier coefficients:
$N_\mathrm{max}$ is the maximum magnitude of the Fourier vectors $\bs{n}$
considered. We show the radial (bottom for all) and vertical action errors
for thin- (blue circles), thick- (green hexagons), halo- (red squares) and
stream-like (purple triangles) orbits. The halo points are labelled by
$N_\mathrm{max}$. For all cases we fix $N_T=24$ and $N_\mathrm{samp}=2400$.}
\label{Fig::Genfunc_converg}
\end{figure}

In Fig.~\ref{Fig::Genfunc_converg} we show the error in the actions for the
four representative orbits as a function of $N_\mathrm{max}$ and computation
time using $N_T=24$ and $N_\mathrm{samp}=2400$. As expected, the errors
generally decrease and the computing time increases as $N_\mathrm{max}$
increases. However, increasing $N_\mathrm{max}$  sometimes {\it increases}
the errors.  For small $N_\mathrm{max}$ the computing time is dominated by
the orbit integration, with the consequence that the time required to compute the
actions of the thin-disc orbit, which is
easy to integrate, is significantly smaller than that required to compute the
actions of the halo orbit. For large $N_\mathrm{max}$, by contrast, the
computing time is dominated by equation solving, and similar times are
required to compute the actions of all four orbits of
Section~\ref{SingleOrbs}.


\subsection{\texttt{tact} Code}

`The Action Computation Tool' or \texttt{tact} is a publicly-available code
implementing all the algorithms given in this paper. It is available at
\href{https://github.com/jls713/tact}{https://github.com/jls713/tact}. We
also provide programs to produce the plots given in this paper, and there is
documentation describing each of the algorithms. The code is written in
C++ but code is provided to compile the routines into a Python
library. The code can work in tandem with the code \texttt{TM} (Torus Mapper), which
is described by \cite{BinneyMcMillan2015}.

Each algorithm discussed in the paper is implemented as an
\texttt{Action\_Finder} class. These classes take as arguments instances of a
potential class (\texttt{Potential\_JS}), and all have methods
\texttt{actions} and \texttt{angles} that return the actions, and angles and
frequencies respectively given some Cartesian 6D phase-space point $(x,y,z,v_x,v_y,v_z)$.
It should be simple for users to specify their own potential by implementing
an inherited class of \texttt{Potential\_JS} and defining a method
\texttt{Phi} that returns the potential and a method \texttt{Forces} that
returns the force. We also provide a simple \texttt{Orbit} class that may be
used with custom potentials.

\section{Conclusions}\label{Sect::Conclusions}

We have reviewed the currently available algorithms for
estimating actions, angles and frequencies in general potentials. We began by presenting a discussion of the separable potentials in which these quantities can be expressed as simple quadratures. The most general such class of potentials is the triaxial St\"ackel potentials of which axisymmetric St\"ackel potentials and spherical potentials are limiting cases. The only case in which the actions can be computed analytically is the isochrone potential. The existence of actions in more general systems has been demonstrated from numerical experiments but there is no way to simply compute the actions in a general potential. The results from the special separable cases have inspired and aided work in creating numerical algorithms for action estimation in general potentials and we have collected together and summarised these algorithms.

We focused
on axisymmetric potentials and for tests used a realistic model of the
potential of our Galaxy. The presented methods fall into two groups: non-convergent
methods and convergent methods.
The convergent methods should produce ever more accurate action estimates
given increased computing resource.
The non-convergent methods all centre on
defining one-dimensional effective potentials that enable actions to be
obtained from a one-dimensional integral. Directly or indirectly they all
derive from St\"ackel potentials, and their accuracy deteriorates with the
distance of the actual potential from the nearest St\"ackel potential.  The
convergent and non-convergent methods are entirely complementary. For some
applications one requires a few accurate action computations (e.g. stream
modelling) whilst for other applications one requires many less accurate
actions (e.g. disc or halo modelling).

We began by comparing the actions estimated along four representative orbits
in a multi-component axisymmetric Galactic potential. We went on to compare
the accuracy of the actions, frequencies and angles for a large number of
orbits in the same potential, and we finally discussed the computational
speed of each method. The time required to compute an action varies by three to four
orders of magnitude between the cheapest method, the St\"ackel Fudge, and the
most expensive method, ItTC.  We gave distribution functions for components of the Milky Way and compared the accuracy of the action estimates yielded from the various methods with the relative error in the corresponding evaluation of the distribution function. The accuracy of
St\"ackel Fudge v1 is sufficient for modelling the smooth phase-mixed components of the Galaxy (the disc and halo) with the possible
exception of some thick-disc orbits. The convergent methods O2GF and
ItTC provide significantly more accuracy than one normally requires,
except possibly at powerful resonances and when modelling cold structures such as tidal streams.

The accuracy of all methods deteriorates in the vicinity of major resonances
such as the resonance $\Omega_r/\Omega_z=1$. Orbits can become trapped at
such resonances, and the actions we seek to evaluate here are not the true
actions of a trapped orbit. Instead one independent action should be a linear
combination of the actions sought here, and the other independent action
quantifies the amplitude of the trapped orbit's libration about an underlying
closed orbit.  The non-convergent methods typically lack the ability to
identify a resonance and the returned actions are at best the actions of a
similar orbit in a nearby integrable Hamiltonian (in which trapping does not
occur). Currently there is no general-purpose tool that will evaluate the
true actions of a trapped orbit, although \cite{KaasalainenB} showed that the
actions and angles of trapped orbits can be evaluated by torus mapping.

Action, angle and frequency variables have proved important tools for modelling dynamical systems, and with the arrival of the Gaia data on the horizon the need for robust, accurate and rapid routines for estimating the actions is great. We have demonstrated that there are powerful algorithms that are appropriate for a wide range of modelling requirements and that promise to be indispensable for understanding and dissecting the Gaia data.

\section*{Acknowledgments}
JLS acknowledges the financial support of the Science and Technologies Facilities Council (STFC). This project was made possible through the use of following open-source software: the Gnu Science Library \citep[GSL]{GSL}, the \texttt{cuba} multidimensional integration library \citep{cuba}, and the Python packages \texttt{numpy} \citep{numpy} and \texttt{matplotlib} \citep{matplotlib}. We thank Angus Williams for reading a draft of this paper and constructing the backronym `The Action Computation Tool' for the \texttt{tact} code.

\bibliography{bibliography}
\bibliographystyle{mn2e}

\appendix
\section{Relative errors in distribution functions for Galactic components}\label{Appendix}
\subsection{Thin and thick discs}

For both the thin and thick discs we use the quasi-isothermal distribution
function presented in \cite{Binney2010}. This distribution function is
nearly-separable in the actions such that the radial and vertical dependence
is given by
\begin{equation}
f_\mathrm{disc}(J_r,J_z) \propto \exp \Big(-\frac{\kappa J_r}{\sigma_R^2}-\frac{\nu J_z}{\sigma_z^2}\Big),
\end{equation}
where $\kappa$ and $\nu$ are the epicyclic frequencies and $\sigma_R$ and
$\sigma_z$ are velocity dispersion parameters. We adopt the parameter values
from \cite{Piffl2014} that were found to produce a good fit to the RAVE data.
From a distribution function of this form the relative error in the
distribution function is given by
\begin{equation}
\Big(\frac{\Delta f_\mathrm{disc}}{f_\mathrm{disc}}\Big)^2 = \sqrt{\Big(\frac{\kappa\Delta J_r}{\sigma_R^2}\Big)^2+\Big(\frac{\nu\Delta J_z}{\sigma_z^2}\Big)^2}.
\end{equation}
Note that due to the form of the distribution function the relative error in
the distribution function is independent of the actions.

\subsection{Stellar halo}

For the stellar halo we use the distribution function from
\cite{WilliamsEvans2015}. This distribution function was fitted to the Blue
Horizontal Branch stars from the SEGUE survey and has the form
\begin{equation}
f_\mathrm{halo}\propto \mathcal{L}^{-q}(\mathcal{L}^2+J_b^2)^{-(p-q)/2},
\end{equation}
where
\begin{equation}
\mathcal{L} = \mathcal{F}D_0 J_r+|J_\phi|+J_z.
\end{equation}
$p=0.83$ and $q=9.16$ govern the inner and outer slopes of the density
profile and $J_b=3600\kpc\kms$ governs the break radius of the density
profile. $D_0=1.52$ produces the isotropic model and $\mathcal{F}=0.59$
controls the anisotropy of the model.

As with the disc distribution functions we use this expression to find the
relative error in the distribution function as a function of the actions.
Here we note that, unlike with the discs, the relative error \emph{does}
depend on the actions. However, we see from Fig.~\ref{MultiOrbits} that the
dependence is relatively weak.

\subsection{Streams}

Streams are formed from material tidally stripped from a progenitor and tend
to form cold thin structures that are well approximated by a Gaussian in
action space. If the progenitor has a velocity dispersion $\sigma_v$ the
widths of the resulting action distributions are well approximated by
\citep{EyreBinney2011}
\begin{equation}
\begin{split}
\sigma_{J_r} &= \frac{\sigma_v(R_a-R_p)}{\pi},\\
\sigma_{J_\phi} &= \frac{\sigma_v R_p}{\pi},\\
\sigma_{J_z} &= \frac{2\sigma_v z_\mathrm{max}}{\pi}
\end{split}
\label{ActionSpread}
\end{equation}
where $R_p$ is the pericentric radius of the progenitor orbit, $R_a$ the
apocentric radius and $z_\mathrm{max}$ the maximum height above the Galactic
plane.

\cite{Bovy2014} and \cite{Sanders2014} demonstrate that a simple, but
realistic stream model can be constructed in angle-frequency space.
Therefore, it is of interest to know the error required to resolve the
angle-frequency structure of the stream. The frequency spread of the stream
can be related to the action spread via the Hessian matrix as
\begin{equation}
\frac{\partial^2 H}{\partial J_i \partial J_j}\equiv \frac{\partial \Omega_i}{\partial J_j},
\end{equation}
which is computed by finite differencing the frequencies for a series of tori
constructed around the true torus. We use this matrix to convert the action
spreads in equation~\eqref{ActionSpread} into frequency spreads
$\bsigma_{\Omega i}$.

McMillan (2016, in prep.) shows that the initial angle spreads of
particles released into a tidal stream from a progenitor are given by
\begin{equation}
\begin{split}
\sigma_{\theta_R} &= \frac{\pi r_s}{(R_a-R_p)},\\
\sigma_{\theta_\phi} &= \frac{\pi r_s}{R_p},\\
\sigma_{\theta_z} &= \frac{\pi r_s}{2 z_\mathrm{max}},
\end{split}
\end{equation}
where $r_s$ is the scale radius of the stream progenitor. However, the angle
spread in the stream increases linearly in time at a rate governed by the
frequency separation from the progenitor so for most streams we do not need
to resolve the angle spreads in such fine detail.

\label{lastpage}
\end{document}